\documentclass[conference]{IEEEtran}

\usepackage{fancyhdr, epsfig, epsf, amsthm, amsmath, amssymb, amsfonts, subfigure, color}
\usepackage{enumerate}
\usepackage{dsfont}
\usepackage[noadjust]{cite}
\usepackage{stfloats}

\newcommand{\chapternote}[1]{%
 \let\thempfn\relax% Remove footnote number printing mechanism
  \footnotetext[0]{\emph{#1}}% Print footnote text
  }

\begin{document}

%\fontsize{16.5}{24}\selectfont
\title{\fontsize{16.5}{24}\selectfont Massive UAV-to-Ground Communication and\\ its Stable Movement Control: A Mean-Field Approach}
% \title{ Massive UAV-to-Ground Communication and \\its Stable Movement~Control}

\author{\IEEEauthorblockN{ $^\ast$Hyesung Kim, $^{\dagger}$Jihong Park, $^{\dagger}$Mehdi Bennis, and $^\ast$Seong-Lyun Kim}

\IEEEauthorblockA{\\\small
$^\ast$School of Electrical and Electronic Engineering, Yonsei University,
Seoul, Korea, email: \{hskim, slkim\}@ramo.yonsei.ac.kr\\
$^\dagger$Centre for Wireless Communications, University of Oulu, Finland, \{jihong.park, mehdi.bennis\}@oulu.fi
}

%\thanks{H. Kim and S.-L. Kim are with the Radio Resource Management $\&$ Optimization
%Laboratory, Department of Electrical and Electronic Engineering,
%Yonsei University, Seoul, Korea (email: \{hskim, slkim\}@ramo.yonsei.ac.kr).}
%\thanks{$^{\dagger}$J. Park and M. Bennis are with Centre for Wireless Communications, University of Oulu, 4500 Oulu, Finland (email: \{jihong.park, bennis\}@ee.oulu.fi).}
}

\maketitle

\begin{abstract}
This paper proposes a real-time movement control algorithm for massive unmanned aerial vehicles (UAVs) that provide emergency cellular connections in an urban disaster site. While avoiding the inter-UAV collision under temporal wind dynamics, the proposed algorithm minimizes each UAV's energy consumption per unit downlink rate. By means of a mean-field game theoretic flocking approach, the velocity control of each UAV only requires its own location and channel states. Numerical results validate the performance of the algorithm in terms of the number of collisions and energy consumption per data rate, under a realistic 3GPP UAV channel model.
\end{abstract}

%\linespread{2}

\section{Introduction}

Cellular connections in our everyday life are about to be ubiquitously reliable in 5G cellular systems \cite{PetarURLLC:17,MehdiURLLC:18,JParkSpaSWiN:17}. The remaining cellular coverage holes would then come from disaster scenarios, which significantly disrupt the search and rescue operations \cite{Akyildiz:17}. To fill these holes quickly and efficiently, it is envisaged to utilize unmanned aerial vehicles (UAVs) that support air-to-ground cellular communications \cite{dynamic_u_cell_cont_16,uav_ee1,RZhang17,Mozaffari17,Mozaffari16}. In this work, we focus particularly on an urban disaster scenario requiring a large number of emergency connections that are enabled by a massive number of UAVs. 

The major technical challenge is the real-time movement control of the massively deployed UAVs. To elaborate, when the ground users are crowded around a disaster hotspot as shown in Fig.~1, the optimal UAV locations for the air-to-ground communication are also likely to be concentrated. The resulting inter-UAV distances may be too short, leading to the inter-UAV collisions caused by their swaying in the wind. To keep their collision-safe distances stable under time-varying wind dynamics, it is thus necessary to adjust the UAV locations continuously. At the same time, due to the limited battery capacity, UAVs need to maximize their energy efficiency, which is difficult to be optimized \emph{per se} even for a single UAV under a given constant wind velocity \cite{RZhang17}.

In this paper, we tackle this real-time massive UAV control problem by proposing a distributed UAV velocity control algorithm. The key idea is to form a flock of UAVs, i.e., a group of UAVs moving at an identical velocity, thereby avoiding the inter-UAV collisions. At the same time, we aim at maximizing the UAV energy efficiency defined as the consumed energy for both wireless transmissions and mechanical movements per unit downlink rate.

\begin{figure}
\centering
\includegraphics[width=9.2cm]{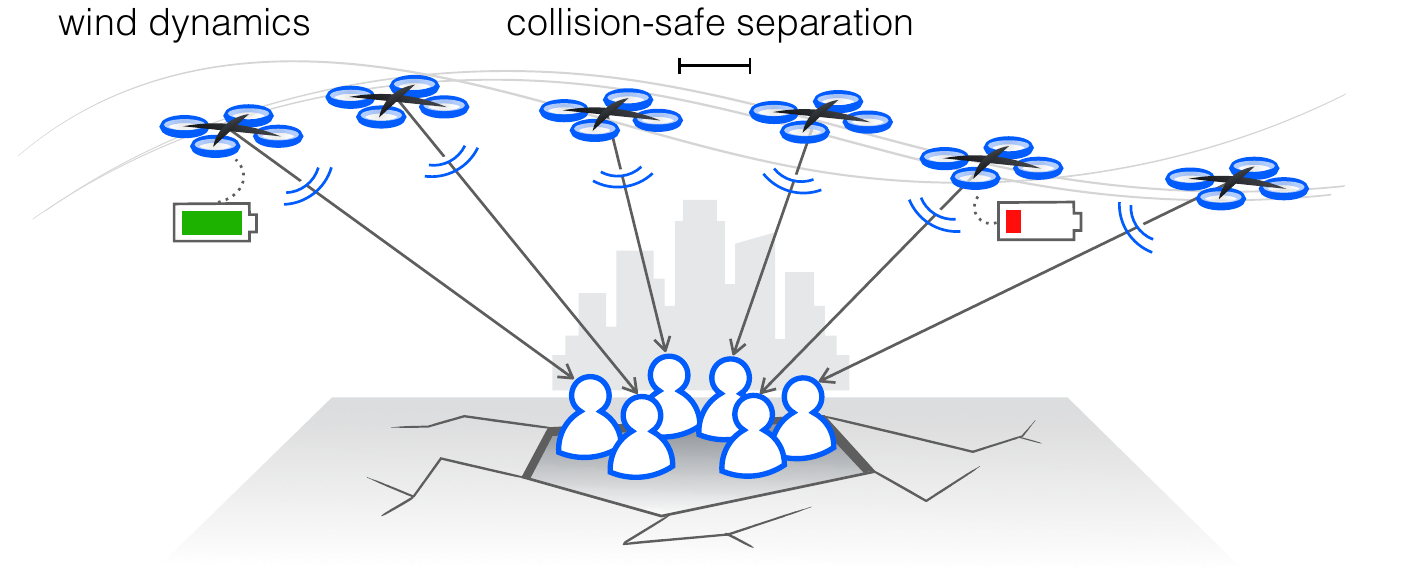}   %height=0.33
\caption{{\small {An illustration of a UAV-to-ground cellular network that supports the ground users crowded around a disaster hotspot.}} } \label{sys_uav} 
\end{figure}

To this end, we propose a mean-field game theoretic UAV flocking algorithm. In this algorithm, each UAV controls the velocity so as to minimize the weighted sum of its long-term energy consumption per downlink rate and its flocking cost. Here, the energy consumption takes into account the amount not only of downlink transmissions but also  mechanical movements. The flocking cost follows from the Cucker-Smale (CS) flocking algorithm that guarantees the velocity convergence within a finite time span \cite{CS_avoidance}. 

When the number of UAVs is large, the velocity control problem of each UAV can be formulated as a mean-field game, i.e., a non-cooperative game between each UAV and a single virtual agent reflecting the ensemble average control decision of all the UAVs.
Then, an individual UAV's velocity is determined by solving a partial differential equation (PDE), known as the Hamilton-Jacobi-Bellman (HJB) equation \cite{exist_HJBsol1}. The resultant UAV movements are then obtained by solving another PDE known as the Fokker-Planck-Kolmogorov (FPK) equation. 
Finally, each UAV can thereby decide its velocity only using its own location and channel states, without exchanging any information with other UAVs. 

The effectiveness of the proposed mean-field flocking algorithm is validated by simulation using the 3GPP air-to-ground channel model \cite{3gpp_ch}. The proposed approach saves up to 55\% average energy consumption per downlink rate, compared to a baseline flocking scheme without considering energy efficiency under the same target collision probability.

\begin{figure*}[hb]
\hrule
\vskip 5pt
\begin{equation}{
L_i(t) = \left\{ \begin{array}{ll}
30.9+(22.25-0.5\log_{10}h)\log_{10}d_z(t)+20\log_{10}f_c,
\text{if LOS link,}\\
\max\{L_i^{LOS},32.4+(43.2-7.6\log_{10}h)\log_{10}d_z(t)+20\log_{10}f_c\},
\text{if NLOS link,} \end{array}\right.} \nonumber\label{pathloss_1}
\end{equation}
where $y_i$ denotes the ground user associated to UAV $i$, $d_z(t)=\Arrowvert z_i(t)-y_i \Arrowvert$ is the inter UAV-user distance, $f_c$ is the carrier frequency, where $L_i^{LOS}$ is the path loss for the LOS link.  The LOS/NLOS link states are stochastically determined by the LOS probability~$P_{LOS}$:
\begin{equation}{\small 
P_{LOS} = \left\{ \begin{array}{ll}\!\!
1, \quad\text{if }\sqrt{d_z^2-h^2}    \leq  d_o,\\
\!\!\ \!  \frac{d_o}{\sqrt{d_z^2-h^2}}+\exp\left\{\left( \frac{-\sqrt{d_z^2-h^2}}{p1}\right ) \!\left( 1-\frac{d_o}{\sqrt{d_z^2-h^2}}  \right) \right\}   ,
\quad \text{if } \sqrt{d_z^2-h^2}  >  d_o,
 \end{array}\right.} \label{pathloss_1}\nonumber
\end{equation}where $\small{d_o\!=\!\max[294.05\log_{10}h\!-\!432.94,18]}$, and $p_1=233.98\log_{10}h-0.95$.  
Note that the NLOS probability is given by ${\small P_{NLOS}=1-P_{LOS}}$, and this model holds for the given altitude $22.5 \text{ m} \leq h \leq 300 \text{ m}$.
\end{figure*}

\textbf{Related work} Trajectory optimization for UAVs has been studied primarily from a robotics/control perspective. 
Most of the current UAV literature is focused on solving air-to-ground wireless communication problems in sparse deployment scenarios \cite{dynamic_u_cell_cont_16,uav_ee1,Mozaffari17,Mozaffari16}.  These include power control, altitude optimization, UAV location, and so forth.  Nevertheless these works overlook specifics of UAV in terms of energy efficiency, collision avoidance and massive UAV deployments.  Fewer works study the interplay between UAV's trajectory (control) and the wireless communication performance such as \cite{RZhang17, Grotli12,Franco15}.
However while interesting these works do not consider the challenging scenario involving a  massive deployment of UAVs. To the best of our knowledge this is the first work  to fill that void.

This article is structured as follows: The system model is described in Section II. The flocking problem formulation and analytical results applying MF game theory are described in Section III. The performance of the proposed algorithm is numerically evaluated in Section IV. Finally, concluding remarks are given in Section V.

% To this end, we adjust UAV velocities in a distributed way as in the Cucker-Smale (CS) flocking algorithm that guarantees the velocity convergence within a finite time span \cite{CS_avoidance}. In the CS algorithm, an individual UAV's velocity is determined by differential equations. The resultant UAV movements are then obtained by the instantaneous locations and velocities of all the UAVs, inducing too frequent information exchanges and recurrent velocity decisions with high complexity. Furthermore, the algorithm is unable to take into account its corresponding energy efficiency.

% as done in \cite{CS_flock1,HKimCachingJSAC:18}

% Also, the system asymptotically achieves the $\epsilon$-Nash equilibrium \cite{MFG_application}. 

\section{System Model}
We consider an air-to-ground downlink network comprising $N$ UAVs at an identical altitude of $h$ meters. The coordinates of the $i$-th UAV at time $t$ is denoted as $z_i(t)\in\mathbb{R}^3$. The UAV location is affected by its own velocity $v_i(t)$ and also by a given wind velocity. Following \cite{WindWiener:16}, we assume the wind dynamics follows an Ornstein-Uhlenbeck process, and thereby represent the temporal dynamics of $z_i(t)$ as:
\begin{equation}
dz_i(t)=(v_i(t)+A)dt+\eta_{A}\text{d}W_i(t), \label{dynamics1}
\end{equation}
where $A$ denotes the average wind velocity, $\eta_{A}>0$ is the wind velocity variance, and $W_i(t)$ is the standard Wiener process, which is identical and independent among UAVs.

$N$ ground users are concentrated at a disaster hotspot, i.e., $h=0$. They are uniformly distributed over a ball centered at the origin with radius~$r$. In order to maximize the safety guarantee, we assume that each user is served by a single dedicated UAV, with the use of disjoint frequency bandwidth $B$. More sophisticated multiple access schemes with reliability guarantees, as done in \cite{PetarURLLC:17,MehdiURLLC:18,ParkGC:18}, are applicable with additional complexity. This is  deferred to future work.

The channel link between each UAV and the associated user follows from the UAV channel model provided by the 3GPP specifications \cite{3gpp_ch}. The path loss is stochastically determined by  line-of-sight (LOS) and non-line-of-sight (NLOS) link states, depending on the height of UAV, the distance from its ground user, and the carrier frequency as described at the bottom of this page.
The instantaneous UAV downlink rate $R_i(t)$  is given by:
\begin{align}
R_i(t)=B\log_2\left( 1+\frac{g_i(t)P_{u}\cdot10^{-L_i(t)/10}}{N_oB}     \right),
\end{align}
where $g_i(t)$ is the fading coefficient, $N_o$ is the noise spectral density, and $L_i(t)$ denotes the path loss.

We use $\Arrowvert\!\cdot\!\Arrowvert$ to indicate the Euclidean norm. The notation
$\nabla f$ denotes the gradient of a function $f$. The operation $\text{inf}$ implies the infimum. The indicator function $\mathds{1}(A)$ returns $1$ if the event $A$ occurs, and 0, otherwise.

% We further assume that each UAV is transmitting data to its
% user with a fixed transmission power of $P_u$, $B_i$
%  is the bandwidth exclusively allocated to UAV $i$, and $f_c$  is the 
% carrier frequency. 

%The SDE \eqref{dynamics1} represents a stochastic fluctuation of request probability around the mean value ?k,j. Due to the randomness of the term Wk,j(t), the request probability xk,j(t) deviates from the long-term mean value ?. If xk,j(t) becomes larger, the term r(?k,j ?xk,j(t)) has a negative value and makes dxk,j(t) head back to the mean value by complementing the

%\subsection{UAV-to-Ground path loss model}
%

%

\section{Energy-Efficient Massive UAV Flocking}

Our primary goal is to control the velocities of a massive number of UAVs in a distributed way, so as to minimize the long-term energy consumption per downlink rate subject to collision-avoidance guarantees. The corresponding problem formulation and our proposed solution are provided in the following subsections.

\subsection{Flocking Problem Formulation}

We define collision as an event when an inter-UAV distance is smaller than a collision-safe separation distance $d_s>0$. The UAV velocity control should abide by the empirical collision probability not exceeding a target collision probability $\epsilon$. For $N$ UAVs flying during $T$ time span, this collision-avoidance constraint is given for UAV $i$ as:
\begin{equation}
\frac{1}{TN}\int_{t=0}^T \sum_{j=1}^N\mathds{1}\left(\Arrowvert z_i(t)-z_j(t)\ \Arrowvert < d_s \right)dt \leq \epsilon. \label{constraint_ca}
\end{equation}

This constraint requires the entire history of all the UAV movements during $T$, which prevents a real-time control. We detour this problem by proposing a UAV flocking algorithm. The key idea comes from the fact that the collision probability of the flocked UAVs with an identical velocity asymptotically converges to $0$, as $T$ increases.

To this end, we adjust UAV velocities in a distributed manner as in the Cucker-Smale (CS) flocking algorithm that guarantees the velocity convergence within a finite time span \cite{CS_avoidance}. Unfortunately, the original CS algorithm requires the instantaneous locations and velocities of all the UAVs, inducing too frequent information exchanges and recurrent velocity decisions with high complexity. Furthermore, the algorithm is unable to take into account its corresponding energy efficiency.

We overcome these limitations by leveraging a non-cooperative game theoretic approach, as done in \cite{CS_flock1,ParkGCEE:16,ParkGCMM:16,KimICC:17,HKimCachingJSAC:18}. In our modified CS flocking algorithm, each UAV tries to minimize its long-term energy cost and its flocking cost. The energy cost $E_i(t)$ of the $i$-th UAV is the total energy consumption for downlink transmissions and mechanical movements per unit downlink rate, given as:
\begin{flalign}
E_i(v_i(t),{z_i(t)})=\frac{e_{m,i}(t)+e_{w,i}(t)}{R_{i}(t)}, \label{EC_function}
\end{flalign} 
where $e_{m,i}(t)=\frac{1}{2} a_m v_i(t)^2$ denotes the energy consumption for movement control, and $e_{w,i}(t)=P_u + a_e$ indicates the energy consumption for transmission. The parameter $a_m$ is the mass of the UAV, $P_u$ is the air-to-ground transmission power, and $a_e$ is a fixed energy consumption independent of $P_u$.

The flocking cost follows from \cite{CS_flock1}. Denoting  $\boldsymbol{v}(t)\!=\!\{\!v_1(t),...,v_N(t)\!  \}$ as the set of UAV velocities and  $\boldsymbol{z}(t)=\{\!z_1(t),...,z_N(t)\! \}$ as the set of their locations at time $t$, the flocking cost $F_i(\boldsymbol{v}(t),\boldsymbol{z}(t))$ is given as:
\begin{flalign}
F_i(\boldsymbol{v}(t),\boldsymbol{z}(t))= \frac{1}{N}\sum_{j=1}^N \frac{\Arrowvert v_j(t)-v_i(t)\Arrowvert^2}{(1/\gamma+\Arrowvert z_j(t)-z_i(t)\Arrowvert^2)^\beta}, \label{flocking_cost}
\end{flalign}
where $\gamma$ is a collision aversion factor and $\beta \leq 0.5$ is a positive constant. 
This cost is an increasing function of the number of collisions, thereby reflecting the left-hand-side of the collision-avoidance constraint \eqref{constraint_ca}. As seen by the numerator of $F_i(\boldsymbol{v}(t),\boldsymbol{z}(t))$, the flocking cost also increases with the relative velocities of UAVs, so as to make the velocities converge to an identical flocking velocity.

%the framework of non-cooperative stochastic differential game and mean-field theory. Furthermore, the reliability to CA constraint \eqref{constraint_ca} can be incorporated into the objective cost function, relaxing the problem with the lower number of constraints.

%
%\begin{flalign}
%\textbf{(P1)}\quad  \min_{v_i(t)}&\quad \frac{1}{T} \int_0^T {EC_i(v_i(t),\boldsymbol{z(t)})}dt \\\nonumber \\
%\text{subject to}&\quad\frac{1}{N}\sum_{j=1}^N\left\Arrowvert \frac{ v_j(t)-v_i(t)}{(1/\gamma+\Arrowvert z_i(t)-z_j(t)\Arrowvert^2)^{\beta}}\right\Arrowvert \leq \theta ,\label{const_0}\\
%&\quad dz_i(t)=(v_i(t)+A_i)dt+\eta_{A_i}\text{d}W_i(t), \quad\label{const_1}
%\end{flalign} 

\begin{figure*}[ht]
\centering
\includegraphics[angle=0, width=0.99\textwidth]{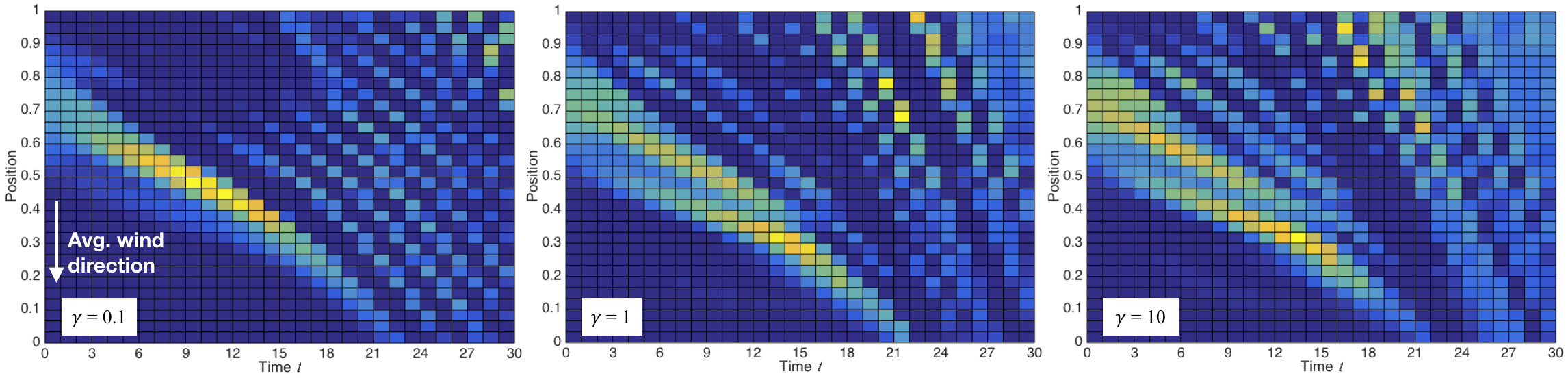} \caption{{ A heat map illustration of the UAV spatial density over time, drawn by using the mean-field distribution $m(z(t))$. For a low collision aversion factor~$\gamma$, e.g., $\gamma=0.1$, each UAV focuses more on its energy efficiency maximization, and follows the wind direction as much as possible. As a result, the UAVs fly in a single large group before they start spreading for the collision avoidance at $t=10$, as shown by the bright-colored group trajectory. For a higher $\gamma$, on the other hand, the UAVs start spreading earlier in order to decrease the collision probability, thereby achieving their flocking faster.
 } }\label{MF_trajectory} 
\end{figure*}

%so that it increases with the relative velocity $\Arrowvert v_i(t)-v_j(t)\Arrowvert$ and decreases with the locational difference $\Arrowvert z_i(t)-z_j(t)\Arrowvert$ with other UAV $j$.
Considering both energy cost and flocking cost, our long-term average (LRA) cost $\mathcal{J}_{i}(t)$ is given by:
\begin{equation}
\mathcal{J}_{i}(t)=\frac{1}{T}\int_{t}^Tw_eE_i(v_i(t),{z_i(t)})+w_fF_i(\boldsymbol{v}(t),\boldsymbol{z}(t))dt, \label{LRA_cost}
\end{equation}
where $w_e$ and $w_f$ are non-negative weight factors. This LRA cost \eqref{LRA_cost} is the objective function of each UAV in the following stochastic differential game:
\begin{equation}
\hspace{-90pt} \textbf{(P1)} \qquad\qquad  \psi_i(t)=\mathop{\text{inf}}\limits_{v_{i}(t)}\text{ } \mathcal{J}_{i}(t),
\end{equation}
\vskip -16pt
\begin{flalign}
\quad \text{subject to }\quad&dz_i(t)=(v_i(t)+A)dt+\eta_{A}\text{d}W_i(t). \quad\label{const_1}
\end{flalign} 

The problem \textbf{P1} has a unique solution that achieves a Nash equilibrium, if 
a joint solution of the following $N$ HJB equations exists \cite{exist_HJBsol1}: 
\begin{align}
0&= \partial_t \psi_i(t)+\mathop{\text{inf}}\limits_{v_{i}(t)} \bigg{[}\underbrace{w_eE_i(v_i(t),{z_i(t)})+w_fF_i(\boldsymbol{v}(t),\boldsymbol{z}(t))}_{(A)}\nonumber \\
 &+\frac{\eta_A^2}{2}\nabla_{z}^2 \psi_i(t)+ \underbrace{(v_i(t)+A}_{(B)} )\nabla_{z_i}\psi_i(t)\bigg{]}. \label{hjb_sdg}
\end{align}\vskip -5pt
\noindent The partial differential equations (PDEs) \eqref{hjb_sdg} are coupled in terms of  mutual distance and  relative velocity. There exists a unique joint solution $\psi^*_{i}(t)$, if the smoothness of the instantaneous cost function (A) and the drift term (B) in \eqref{hjb_sdg} are both guaranteed \cite{exist_HJBsol1}. 
We can demonstrate that this smoothness can verify the existence and uniqueness of the joint solution,  corresponding to the Nash equilibrium of the problem \textbf{P1}. 
%The corresponding instantaneous velocity control of UAV $i$ is given by:
%\begin{align}
%v_i^*=\frac{ \frac{w_f}{N}\sum_{j=1}^N\left[\frac{2v_j(t)}{(1/\gamma+\Arrowvert z_i(t)-z_j(t)\Arrowvert^2)^{\beta}}\right] -\nabla_{z_i}\psi_i(t)        }{      \frac{a_ew_e}{R_{i}(z_i(t))}+\frac{w_f}{N}\sum_{j=1}^N  \frac{2}{(1/\gamma+\Arrowvert z_i(t)-z_j(t)\Arrowvert^2)^{\beta}}            },\label{v_sdg}
%\end{align}
%Unfortunately, the

Unfortunately, solving the problem \textbf{P1} with $N$ HJB equations \eqref{hjb_sdg} is of high complexity and, furthermore, requires the exchange of instantaneous velocities and positions of all other UAVs. In order to overcome these, we leverage a mean-field (MF) game framework described in the following subsection.
For a sufficiently large number of UAVs, this problem can be formulated as a mean-field game, which asymptotically achieves the $\epsilon$-Nash equilibrium~\cite{MFG_application}. Each UAV can thereby decide the velocity using its own location and channel states, without exchanging any information with other UAVs.

\subsection{Mean-field Game Theoretic Flocking Design}

When the number of UAVs $N$ is large, the problem \textbf{P1} is equivalent to a MF game. Thus, we obtain the distribution of other UAVs' positions and velocity controls. Moreover, the expected flocking cost $\bar F(t)$ is written as follows:
\small\begin{equation}
{\bar F_i(v_i(t),z_i(t),m(z(t))) = \!\int_z \frac{m(z(t))\Arrowvert v(z(t))-v_i(z_i(t))\Arrowvert^2}{(1/\gamma+\Arrowvert z(t)-z_i(t)\Arrowvert^2)^\beta}  dz,}\label{F_bar}
\end{equation}\normalsize
where $m(z(t))$ is a resultant UAV-position distribution that corresponds to the individual UAV's velocity control from the HJB equation \eqref{hjb_sdg}. This distribution is called MF distribution \cite{MFG_application}, which is 
 a solution of the following PDE Fokker-Planck-Kolmogorov (FPK) equation given by:
\begin{align}
0= \partial_t m(z(t))+ {(v(t)+A})\nabla_{z}m(z(t))-\frac{\eta_A^2}{2}\nabla_{z}^2 m(z(t)) \label{fpk_1},
\end{align}\vskip -5pt\noindent
This FPK equation \eqref{fpk_1} is coupled with the HJB equation \eqref{hjb_sdg} in terms of the velocity control.
 Solving the HJB equation gives us an individual UAV's velocity control, and the FPK equation stochastically provides its corresponding resultant UAV position distribution. 
 Due to this, the UAV index $i$ is dropped in the FPK equation \eqref{fpk_1}.
 The MF distribution $m(z(t))$ is derived from the empirical distribution $M_t(z(t))=\frac{1}{N}\sum_N \mathds{1}(z(t))$. When $N$ goes to infinity, $M_t(z(t))$ converges to $m(z(t))$. This approach is referred to as the MF approximation  and enables us to achieve the $\epsilon$-Nash equilibrium \cite{EV_JSAC, HKimCachingJSAC:18}.

Let us define $v^*(t)$ as the optimal velocity achieving the equilibrium, which is given by the following Proposition. 
\vskip 10pt
 \noindent {\bf Proposition 1.} {\it The optimal velocity is given by: }
\begin{align}
v_i^*=\frac{ w_f\int_z \frac{2m^*(z(t))v(z(t))}{(1/\gamma+\Arrowvert z_i(t)-z(t)\Arrowvert^2)^{\beta}}dz -\nabla_{z_i}\psi_i(t)        }{      \frac{a_mw_e}{R_{i}(z_i(t))}+w_f\int_z  \frac{2m^*(z(t))}{(1/\gamma+\Arrowvert z_i(t)-z(t)\Arrowvert^2)^{\beta}}dz  },
\end{align} {\it where $m^*(z(t))$ and $\psi^*(t)$ are the unique solutions of  the FPK \eqref{fpk_1} and the following modified HJB equation \eqref{hjb_mfg}, respectively. 
\begin{align}
0&= \partial_t \psi_i(t)+\inf\limits_{v_i(t)} \bigg{[}w_eE_i(v_i(t),{z_i(t)})+\frac{\eta^2}{2}\nabla_{z_i}^2 \psi_i(t)\nonumber \\
 &+w_f\bar F(v_i(t),z_i(t),m\!^*\!(z_i(t)))+ {(v_i(t)+A_i} )\nabla_{z_i}\psi_i(t)\bigg{]}. \label{hjb_mfg}
\end{align}}\vskip 10pt
\noindent {\it Proof}: The optimal $v^*(t)$ is the  
minimizer of the infimum term of the HJB equation \eqref{hjb_mfg}, which is obtained via substituting the expected flocking cost \eqref{F_bar} into the original HJB \eqref{hjb_sdg}. Since the infimum term is convex, we can get straightforwardly $v^*(t)$ from the first order derivative \cite{MF_ref2,MFG_application}. \hfill$\blacksquare$
 \vskip 10pt

  \begin{figure}
\centering
\includegraphics[angle=0, height=0.34\textwidth]{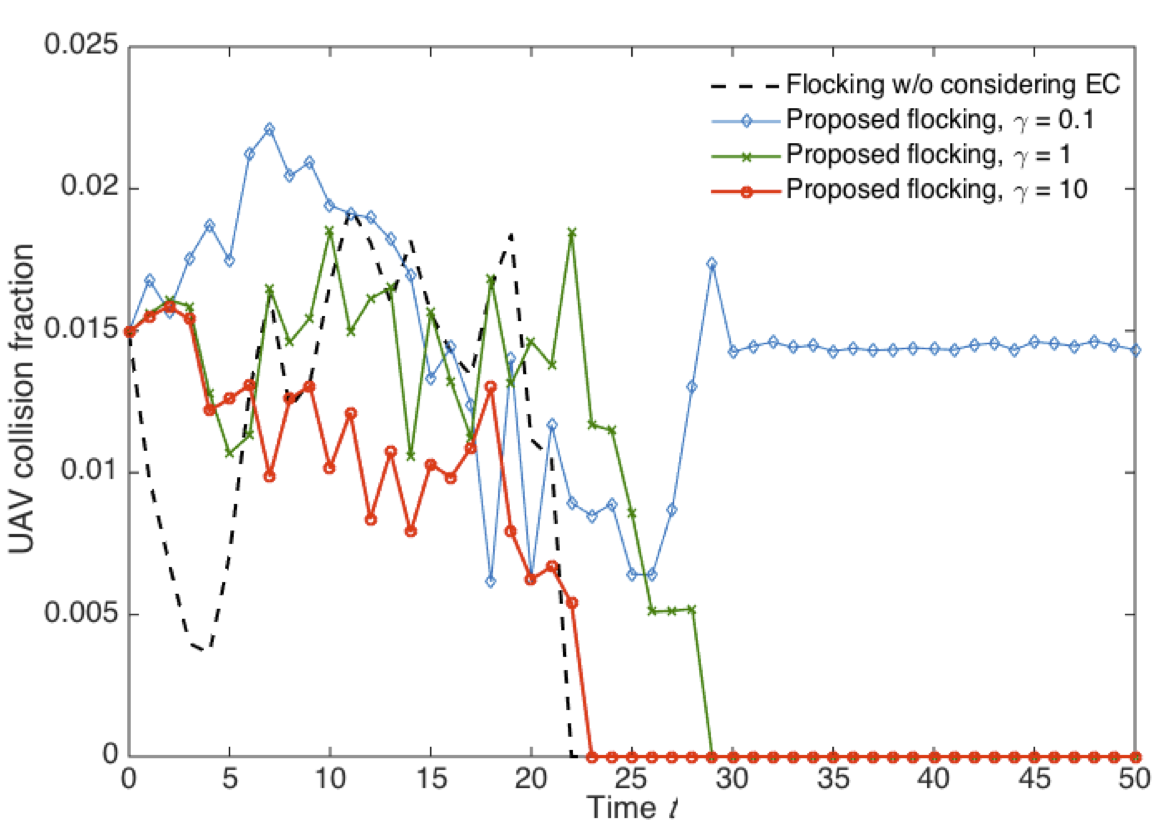} \caption{{\small UAV collision fraction over time.} }\label{fig_collision} 
\end{figure}
\begin{figure}
\centering
\includegraphics[angle=0, height=0.34\textwidth]{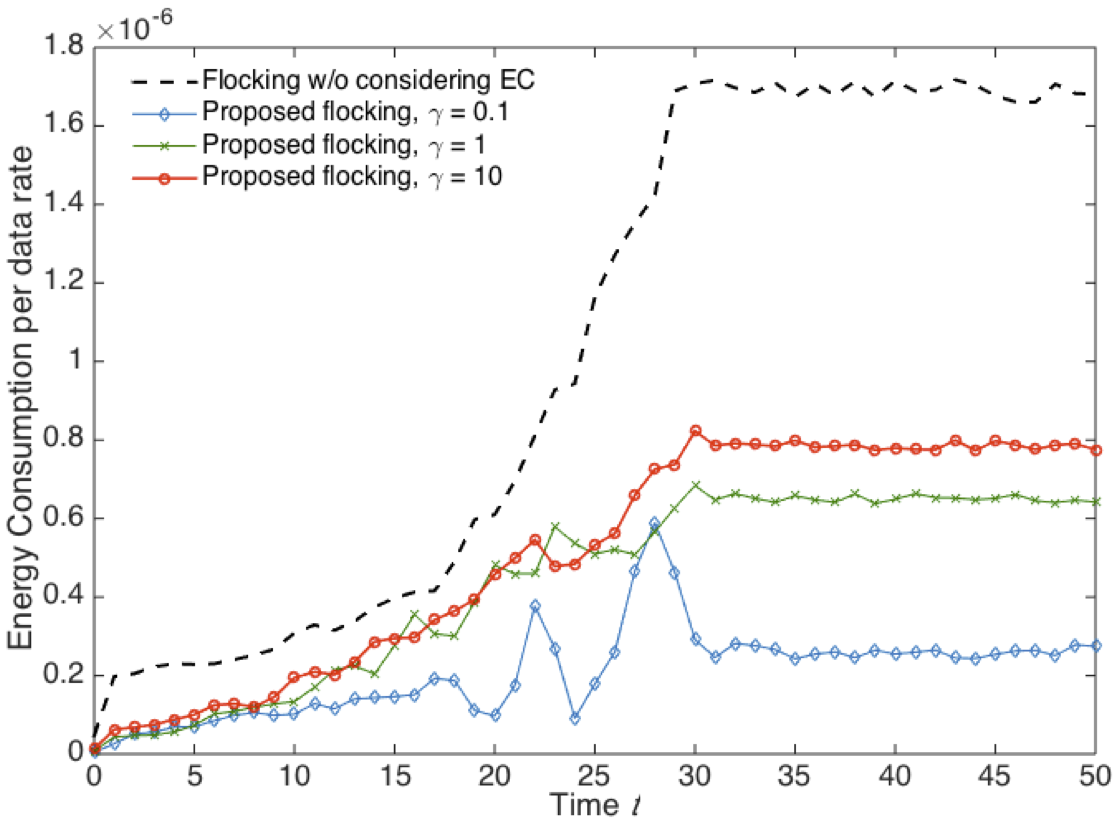} \caption{{\small Energy consumption per data rate over time.} }\label{fig_EC} 
\end{figure}

\section{Numerical Results}

In this section, we present numerical results  to evaluate the proposed algorithm  under temporal wind dynamics \eqref{dynamics1}. 
We assume that one hundred  UAVs are located at the height of 300 m and flying over a line whose length is 300 m. 
The initial UAV distribution of the UAVs $m(z(0))$ is given as a normal distribution 
$\mathcal{N}\!\thicksim\!(210,\sqrt{30})$. 
 Their associated ground users are uniformly distributed on a line of which length is 60m and its center is vertically same to the center of the line space where UAV are flying on. We utilize the urban-micro UAV channel model provided by the 3GPP \cite{3gpp_ch} with
$P_u= 23$ dBm, $f_c= 2$ GHz, $B=20/N$ MHz, $N_o=-173$ dbm/Hz, the shadow fading standard deviation $8$ dB, $A=-3$ m/s, $w_e=1, w_f=1, \eta_A=0.1$, $a_e=0$, $a_m=1, \beta=0.5, d_s=2.5$ m.
In order to
solve the coupled FPK \eqref{fpk_1} and HJB \eqref{hjb_mfg} PDEs using a finite element method, we used the MATLAB PDE solver.

Fig. \ref{MF_trajectory} shows the trajectory of UAVs over time with respect to the different collision aversion factor $\gamma$. The bright-colored point means that there are more UAVs in that area.
For all $\gamma$, in the initial period, most  UAVs fly close together, complying with the wind direction toward the  disaster hotspot. This minimizes the energy consumption per downlink rate by decreasing the mechanical mobility control energy and enhancing the downlink rate for ground users. However, it leads to high collision probability, so that UAVs actively start to be apart from each other against the direction of wind dynamics and aim at ensuring the safety distance $d_s$. Gradually, UAVs are located where  the number of collisions is minimized while keeping the safety distance.

 As the collision aversion factor $\gamma$ increases, the  period of the compliance to wind dynamics becomes shorter and UAVs  promptly adjust their velocity against the wind direction in order to fly far apart from each other. Consequently, UAVs move swiftly where the number of collision is minimized. It is worth noting that energy consumption per downlink rate becomes high in the collision-free state. This trade-off is verified in Fig. \ref{fig_collision} and Fig. \ref{fig_EC}, which show an instantaneous UAV collision fraction, defined as $\frac{1}{N} \sum_{j=1}^N\mathds{1}\left(\Arrowvert z_i(t)-z_j(t)\ \Arrowvert < d_s \right),$ and the energy consumption per downlink rate, respectively. 
 For a higher  value of $\gamma$, the collision fraction converges faster to zero and ensures larger safety distance. However, mechanical energy consumption and distance from the disaster hotspot both increase, yielding higher energy consumption per downlink rate. This means that UAVs determine their velocity against the wind dynamics and increase the inter-UAV distance. Hence, this trade-off can be solved by determining a desirable value of~$\gamma$.

 We also compare our MF flocking algorithm to a flocking algorithm with $w_e=0$, which ignores the energy consumption per downlink rate. As shown in Fig. \ref{fig_collision}, our algorithm with $\gamma=10$ reaches a collision-free regime within the same elapsed time of flocking without considering energy consumption. Furthermore, Fig. \ref{fig_EC} verifies that the proposed MF flocking algorithm saves averagely 55\% of energy consumption compared to the baseline algorithm under the same target reliability to avoid collision.

 \section{Conclusion}
 
In this paper, we proposed an instantaneous movement control algorithm for massive unmanned aerial vehicles (UAVs) providing emergency connections in an urban disaster situation. Our algorithm minimizes the energy consumption per  downlink rate, while avoiding inter-UAV collision under a temporal wind dynamics. 
Leveraging a mean-field game theoretic flocking approach, the control of each UAV only requires its own location and channel states, enabling a fully-distributed control operations. Numerical results validate the performance of our algorithm in terms of the collision fraction and energy consumption per data rate under a realistic 3GPP UAV channel model.
 
 \section*{Acknowledgement}

This research was supported in part by Basic Science Research Program through the National Research Foundation of Korea(NRF) funded by the Ministry of Science and ICT(NRF-2017R1A2A2A05069810), in part by the Academy of Finland project CARMA, in part by the INFOTECH project NOOR, and in part by the Kvantum Institute strategic project SAFARI.

%When the risk aversion parameter $\gamma$ is equal to 0.1, the normalized number of collision does not converge to zero.

%
%
%
%
%
%In this section, we present numerical results  to evaluate the proposed algorithm  under spatio-temporal content popularity and network dynamics as illustrated in Fig.~\ref{system_model_CRP}. 
%Let us assume that the initial distribution of the SBSs $m_0$ is given as a normal distribution. We assume that the storage size $Q(t)$ belongs to a set $[0,1]$ for all time $t$. Assuming Rayleigh fading with mean one, we set the parameters as shown in Table I.
%In order to
%solve the coupled PDEs (the first step of the Algorithm 1) using a finite element method, we used the MATLAB PDE solver.
%
%
%
% \ifCLASSOPTIONcaptionsoff
%   \newpage
% \fi

\bibliographystyle{ieeetr}  % appearance order
\bibliography{IEEEabrv,Massive_U2G}

\end{document}